\documentclass[11pt]{article}
\usepackage{bm,amsfonts,amssymb,amsthm,amsmath,gensymb}
\usepackage{mathrsfs}
\usepackage{graphicx,color}
\usepackage{CJK}
\usepackage{indentfirst,enumerate}

\newcommand{\md}{\mathrm{d}}

\definecolor{orange}{rgb}{1,0.5,0}
\definecolor{rb}{rgb}{1,0,1}
\allowdisplaybreaks[1]

\begin{document}
\title{Computing optimal interfacial structure of ordered phases}
\author{Jie Xu$^1$, Chu Wang$^{2}$, An-Chang Shi$^3$ and Pingwen Zhang$^1$\footnote{Corresponding author}\vspace{12pt}\\
\small
$^1$LMAM \& School of Mathematical Sciences, Peking University, Beijing 100871, China\\
\small
$^2$Program in Applied and Computational Mathematics, Princeton University, \\
\small Princeton, New Jersey 08544, USA\\
\small
$^3$Department of Physics and Astronomy, McMaster University, \\
\small Hamilton, Ontario L8S4M1, Canada\\
\small
Email: rxj\_2004@126.com, chuw@math.princeton.edu, shi@mcmaster.ca, pzhang@pku.edu.cn}
\date{\today}

\maketitle

\begin{abstract}
We propose a general framework of computing interfacial structure. 
If ordered phases are involved, the interfacial structure can be obtained 
by simply minimizing the free energy with compatible boundary conditions. 
The framework is applied to Landau-Brazovskii model and works efficiently. 

\end{abstract}

\section{Introduction\label{Intro}}
Interfaces are transition regions between two different materials, 
two different phases of the same material, 
or two grains of the same phase with different orientations (grain boundaries). 
Being less stable, 
interfacial regions are where most of the molecular motion takes place. 
Therefore, the morphology of interfaces greatly affects the mechanical, 
thermal and electrical properties of the material. Interfaces are frequently 
encountered as planar defects and in phase transitions. 
In specific, the strength and conductivity of the material depend hugely on the 
distribution and morphology of grain boundaries. 
And in first-order phase transitions, the interfacial properties 
play an important role in the nucleation-growth process. 

In the framework of mean-field theory, 
the majority of literatures on interfaces are dedicated to grain boundaries, 
among which, two distinct approaches are usually adopted. 
One approach, which could be summarized as the dynamical approach, 
treats the interface as a transient state and focuses on its dynamics. 
It has been adopted to examine grain boundaries of lamellar and 
cylindral phases \cite{PRE2004,Macro2005,PhTrRSocA2011}. 
The other regards the interface as a metastable state and its morphology 
is considered as a local minimizer of system's energy under some conditions. 
It has been adopted to investigate a few tilted grain boundaries of 
the lamellar phase \cite{PRL1997,JChemP1997,PRE2000,JChP2002} 
and the bcc phase \cite{PRE2009}, 
and twist grain boundaries of several cubic phases \cite{JChemP2009}. 
The dynamical approach enables us to see the dynamical evolution of the 
interface, and is especially useful when we are interested in 
how several grains interact with each other. 
But it is difficult to reach full relaxation. Even if only partial relaxation 
is required, the simulation is more time-consuming because of the 
stability condition. 
The minimization approach, on the other hand, is able to reach full relaxation 
and reveal clearly the structure. 
Furthermore, fast optimization algorithms can be used in the computation. 
Since we aim to analyze the structural properties of interfaces, 
the minimization approach appears the better way. 

Because the interface is away from equilibrium, a delicate handling 
of the boundary conditions is required. 
If the boundary conditions are incompatible with each grain, 
a large cell is needed to reduce the boundary effect, 
for the incompatibility might destroy the structure inside the grain. 
Despite the possibility of only finding inaccurate or even biased structures, 
incompatible boundary conditions are adopted in most relevant works. 
In addition, they may generate several interfaces 
(Fig. 5.20 in \cite{Fredrickson} gives a typical morphology) 
or may not work at all (see the examples in \cite{Macro2005}). 
From another perspective, compatible boundary conditions can sufficiently 
reduce the size of the computational cell. 
The compatibility has been noted in the study of kink \cite{JChemP1997} and 
T-junctions \cite{JChP2002} in lamellar grain boundaries, 
where the basis functions are carefully chosen to retain symmetry properties. 
However, the compatibility is only partially reached, and the choice of 
basis functions cannot be easily extended to other structures. Thus it is 
desirable to propose boundary conditions with universal compatibility. 

In contrast to grain boundaries, few theoretical works report 
the morphology of interfaces between different phases. 
Some works construct the interface profile by a mixing ansatz, 
which is the direct weighted combination of two bulk phase profiles:
\begin{equation}\label{connection}
  \phi(x)=(1-\alpha(x))\phi_1(x)+\alpha(x)\phi_2(x)
\end{equation}
where $\phi_k(x)$ are profiles of two bulk phases, and $\alpha(x)$ is a 
smooth monotone function satisfying $\alpha(-\infty)=0,\ \alpha(+\infty)=1$. 
This method proves to be convenient and effective in several literatures 
\cite{JPSJ2007,SoftM2011}. 
But this artificial approach may exclude the possibility of complex interfacial 
structures as we will present later. 
To our best knowledge, only in \cite{PRL1997} the morphology of a few 
lamellar-cylindral interfaces is reported. 

When considering interfaces between different phases, 
there is another question to care about: 
can we pose it purely as an energy minimization problem? 
Intuitively, the phase with higher energy density will gradually be taken over 
by the other phase. We may try to choose parameters to equalize the energy 
densities, but it is difficult to realize in computation. 
Thus it seems that constraints are needed to prevent the interface from 
moving continuously towards the less stable phase. 
However, we will show in this paper that constraints of such kind 
are not necessary if the bulk phases are ordered. 
The interface will be stuck at a locally optimized position even if 
the energy densities of the two phases are slightly different.
Thus we may let the interface freely relax itself during the computation 
instead of intervening in the process artificially. 

To sum up, we are going to propose a general framework for the computation of 
interfacial structures. 
In this framework, optimization approach is adopted to inherit the advantage 
of full relaxation of the system and fast optimization methods. 
The boundary conditions, or equivalently the choice of function spaces, 
are compatible with two bulk phases. 
The setting is well-posed for ordered phases with energy difference. 
We will apply this framework to Landau-Brazovskii model to illustrate the above 
features. The paper is organized as follows. 
In Sec. \ref{alg}, the general framework of interface problem is described, 
and the well-posedness of the setup is illustrated. Some interfacial structures 
in Landau-Brazovskii model are presented in Sec. \ref{result}. 
Finally we summarize the paper in Sec. \ref{concl}. 
Some details of numerical method are written in Appendix. 

\section{General framework\label{alg}}
\subsection{Boundary compatibility\label{extn}}
When considering an interface between two ordered structures, 
their spatial and orientational relations are essential variables. 
Let us put the two phases in two half-spaces separated by the plane $x=0$. 
Denote the two phases by $\alpha$ and $\beta$ respectively. 
Both phases can be rotated or shifted, represented by orthogonal 
matrices $\mathcal{R}_{\alpha},\ \mathcal{R}_{\beta}$, 
and vectors ${\bm d}_{\alpha},\ {\bm d}_{\beta}$. 
Each of these four quantities has three degrees of freedom. 
Note that their relative position remains unchanged if they are rotated 
together round the $x$-axis or shifted together in the $y$-$z$ plane, 
so there are nine independent degrees of freedom in total. 

Assume that the density profile of a phase can be expressed by 
a single function $\phi$. 
More specifically, we assume that the phase is periodic. 
In this case, the profile is given by the value of $\phi$ in a unit cell, 
which can be written as 
\begin{equation}
  \phi(\bm{r})=\sum_{\bm{k}\in\mathbb{Z}^3}\phi_{\bm{k}}
  \exp\left(i\sum_{j=1}^3 k_j\bm{b}_j\cdot {\bm r}\right). \label{prof_per}
\end{equation}
If the phase is rotated by $\mathcal{R}$, then shifted by $\bm{d}$, the profile 
becomes
\begin{equation}
  \phi(\bm{r};\mathcal{R},\bm{d})=
  \phi(\mathcal{R}^T(\bm{r}-\bm{d}))=\sum_{\bm{k}\in\mathbb{Z}^3}
  \phi_{\bm{k}}\lambda_{\bm{k}}
  \exp\left(i\sum_{j=1}^3 k_j(\mathcal{R}\bm{b}_j)\cdot {\bm r}\right), 
\end{equation}
where $\lambda_{\bm{k}}=\exp(-i\sum_{j=1}^3 k_j\bm{d}^T\mathcal{R}\bm{b}_j)$. 

Now let us explain what is compatibility. 
Suppose that two phases $\alpha$ and $\beta$ are identical. In this case, 
the whole space is filled with the phase $\alpha$ and no interface exists. 
The boundary conditions we propose shall not damage the structure of $\alpha$ 
and generate an interface artificially. 
In other words, the boundary conditions shall be met by the profile 
$\phi_{\alpha}$ as well as $\phi_{\beta}$. 
Equivalently, from the view of function space, it requires that the basis 
functions we choose shall contain that of $\phi_{\alpha}$ and $\phi_{\beta}$. 

We apply the compatibility to the $y$- and $z$-directions. 
Consider the limitation of $\phi$ onto the plane $x=x_0$. 
Denote ${\bm r}'=(y,z)$. Then we have 
$$
\phi_{\alpha}(x_0,\bm{r}';\mathcal{R}_{\alpha},\bm{d}_{\alpha})=
\sum_{\bm{k}\in \mathbb{Z}^3}\phi_{\alpha\bm{k}}\tilde{\lambda}_{\alpha\bm{k}}\exp\left(i 
\sum_{j=1}^{3}k_j\bm{b}_{\alpha j}'(\mathcal{R}_{\alpha})\cdot \bm{r}'\right), 
$$
where ${\bm b}_{\alpha j}'(\mathcal{R}_{\alpha})$ denotes the $y$ and $z$ 
components of $\mathcal{R}_{\alpha}{\bm b}_{\alpha j}$, and
$\tilde{\lambda}_{\alpha\bm{k}}=\lambda_{\alpha\bm{k}}\exp\big(ix_0\sum_{j=1}^3
k_j(\mathcal{R}_{\alpha}\bm{b}_{\alpha j}')_1\big)$. 
It becomes a quasiperiodic function in the plane. 
And for $\phi_{\beta}$, we have 
$$
\phi_{\beta}(x_0,\bm{r}';\mathcal{R}_{\beta},\bm{d}_{\beta})=
\sum_{\bm{k}\in \mathbb{Z}^3}\phi_{\beta\bm{k}}\tilde{\lambda}_{\beta\bm{k}}\exp\left(i 
\sum_{j=1}^{3}k_j\bm{b}_{\beta j}'(\mathcal{R}_{\beta})\cdot \bm{r}'\right). 
$$
Then a natural choice will be the set of quasiperiodic function
\begin{equation}
\mathcal{F}=\left\{f({\bm r}'):f({\bm r}')=\sum_{{\bm k},\bm{l}\in\mathbb{Z}^{3}}f_{{\bm k}}
\exp\left(i\sum_{j=1}^{3} \big(k_j{\bm b}_{\alpha j}'(\mathcal{R}_{\alpha})+l_j{\bm b}_{\beta j}'(\mathcal{R}_{\beta})\big)\cdot {\bm r}'\right) \right\}. 
\label{basis}
\end{equation}
Actually, this choice is also valid if $\phi$ is quasiperiodic, 
namely to substitute the sum over $1\le j\le 3$ with $1\le j\le m$. 
In special cases where 
$k_j{\bm b}_{\alpha j}'(\mathcal{R}_{\alpha})+l_j{\bm b}_{\beta j}'(\mathcal{R}_{\beta})$ 
lie on a 2D lattice, $\mathcal{F}$ is reduced to a set of periodic functions, 
indicating that two phases, with their relative position determined, have public
period in the $y$-$z$ plane. The current work will focus on these special cases.

\begin{figure}
  \centering
  \includegraphics[width=0.7\textwidth,keepaspectratio]{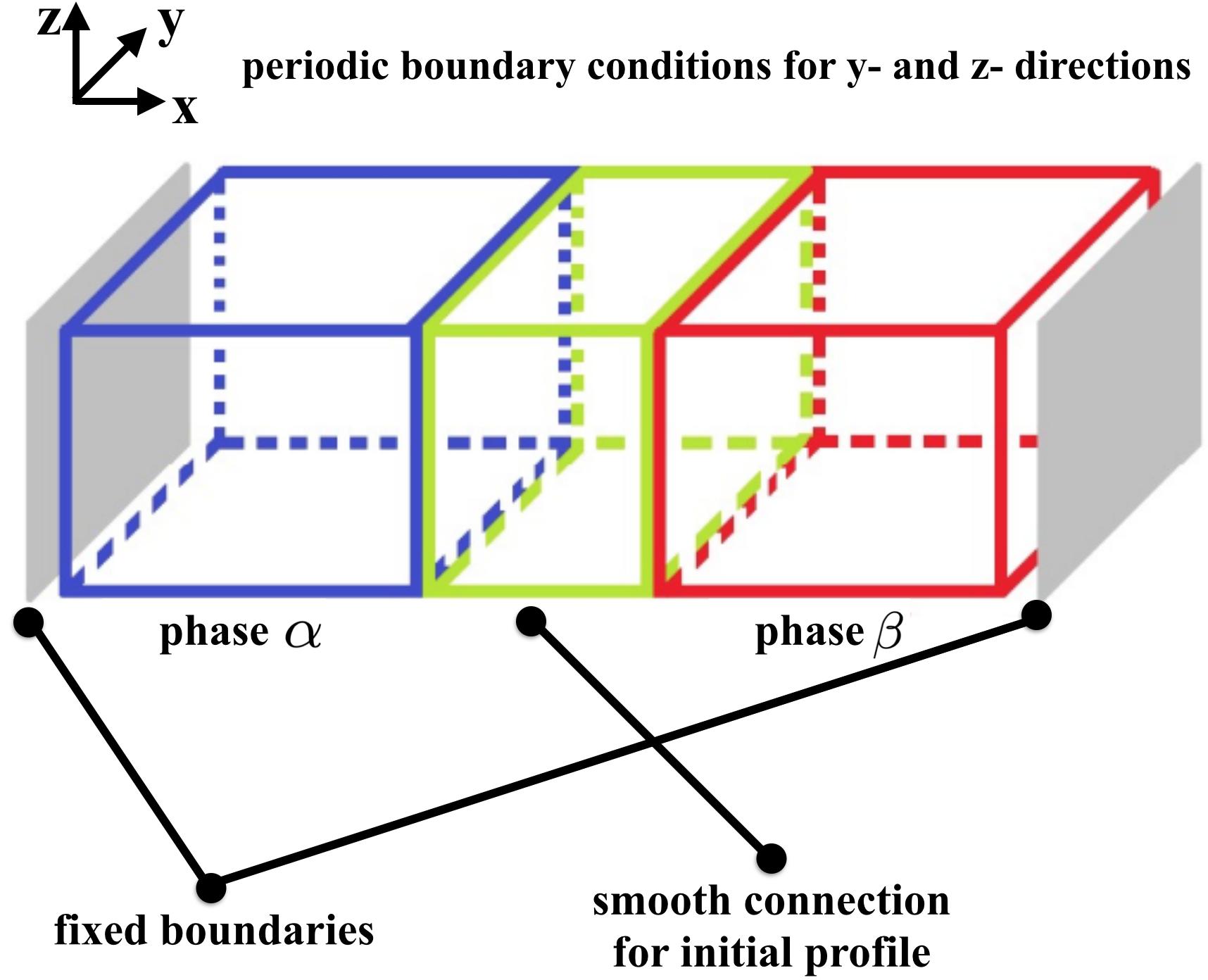}
  \caption{Schematic of the setting of interface problem. }\label{setting}
\end{figure}
In the $x$-direction, we select a length $L$ and set $\phi$ outside 
$[-L,L]$ equal to the bulk value. 
Here $L$ is large enough to contain the transition region. 
Such setting will induce some anchoring conditions at $x=\pm L$ 
dependent on the energy functional. If Laudau-type energy functional is 
used, these conditions can usually be determined by smoothness requirements 
of the density profile $\phi$. For example, if $\phi$ is $C^k$, 
then $\phi,\ldots,\nabla^k \phi$ shall be fixed to bulk values at $x=\pm L$. \
It should be noted that in some other models 
boundary conditions are not directly imposed on the density profile. 
For example, in self-consistent field theory (we refer to \cite{Fredrickson} 
for details), the profile is calculated through a propagator $q$, 
on which boundary conditions are imposed. 
In this case the anchoring conditions can be used on $q$. 

To initialize the density profile for computation in our frame work, 
we use the setting in Fig. \ref{setting}.
We first choose a common period in $y$- and $z$- direction for 
phase $\alpha$ and $\beta$, 
and then fill in the bulk profiles and anchor both ends of the region.
To obtain a smooth initial value, 
the density profile in the middle region is set as the 
convex combination of the bulk densities as in (\ref{connection}). 

\subsection{Existence of local minima}
\begin{figure}
  \centering
  \includegraphics[width=0.5\textwidth,keepaspectratio]{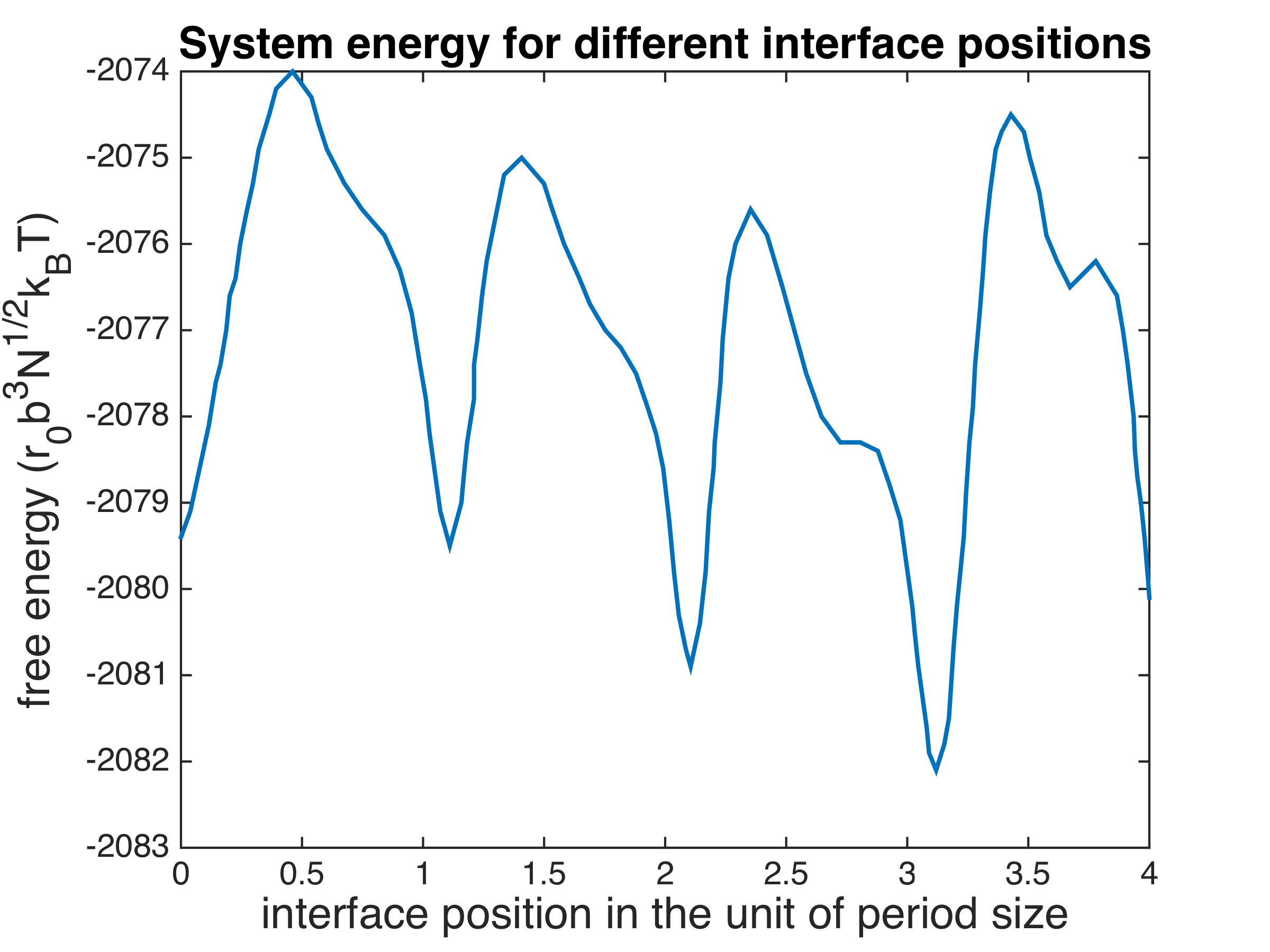}
  \caption{Energy profile in the movement of cylindral-gyroid interface within 
    a single period. The location of interface is calculated using a modified 
    functional distance between the interfacial density and 
    the two bulk densities \cite{str0,simpstr}. 
    The distance is measured by the unit cell of gyroid, 
    which is rescaled to four (the upper bound of the $x$-axis). }
  \label{mep}
\end{figure}
Before solving the optimization problem, 
the well-posedness of this setting should be discussed.
More specifically, we need to demonstrate the existence of local minima 
in our setting. 
It is obvious that for disorder-disorder interface, 
if there exists a difference in the bulk energy densities, 
no matter how small it is, 
the interface would move continuously to the one with higher energy density. 
In this case, the total energy can be written as 
\begin{equation}\label{eng0}
E=f_{\alpha}V_{\alpha}+f_{\beta}V_{\beta}+\gamma S 
\end{equation}
where $f$ is the free energy density, $V$ is the volume of each phase, 
$S$ is the area of interface, and $\gamma$ is the interfacial energy density. 
The isotropy of two phases along the $x$-direction makes $\gamma$ independent 
of interface location. Suppose $f_{\alpha}<f_{\beta}$, 
then $E$ is monotone decreasing when $V_{\alpha}$ increases, 
driving the interface to the phase $\beta$. 
In some earlier works on liquid-vapor nucleation, constraint 
method has been used to fix the location of interface 
\cite{JChP1994,JChP1996,JChP2001}. 
The constraint acts equivalently as the chemical potential that fixes 
the volume fraction of each phase. 
Nevertheless, it is difficult to propose a constraint in 
ordered systems physically meaningful. 

Fortunately, we can benefit from anisotropy in ordered phases. 
The anisotropy implies that $\gamma$ is no longer constant. 
Intuitively we write $\gamma=\gamma(x)$ as a function of interface location, 
and still write the total energy as (\ref{eng0}). 
If $\gamma(x)$ varies more drastically than the bulk energy difference, 
some local minima exist, towards which we may let the interface relax. 
Such condition is easier to be attained when 
the energy difference between two phases is small. This can be achieved by 
choosing model parameters near (but not necessarily on) the binodal line. 
In fact, only when the parameters lie in this range could the interfaces exist. 
Otherwise the phase of higher energy density will disappear under tiny 
perturbation, 
or even become unstable and quickly decompose into the stable phase. 

We would like to give an example to illustrate the how the total energy varies 
as the interface moves. 
We compute the cylindral-gyroid interfaces in the Landau-Brazovskii model 
(\ref{LB}) with parameters $\xi^2=1.0$, $\tau=-0.3$, $\gamma=0.383$. 
Four local minima are found within a single period along the $x$-direction, 
and they are connected with the minimum energy path computed using string 
method \cite{str0,simpstr}. Fig. \ref{mep} shows clearly that the energy 
decreases in a wavy manner as the interface moves. 
The morphology of interfaces at local minima is identical to what is drawn in 
Fig. \ref{match} (in which we draw the interfaces within two periods), 
although we choose different parameters. 

As we mentioned in Sec. \ref{Intro}, quite few works examine interfaces between 
different phases. The reason could be the lack of knowledge of the existence of
local minima produced by microstructures. We feel that the clarification of the 
energy profile would be helpful to the computation of interfacial structure. 

\section{Application to Landau-Brazovskii model\label{result}}
In this section, we apply our framework for interface to 
Landau-Brazovskii (LB) model. We first present 
the cylindral-gyroid and lamellar-gyroid interfaces 
in epitaxially matching cases, in which the local minima are shown clearly. 
Furthermore, a few novel examples of non-matching cases are given 
to show the effectiveness of our framework and algorithm. 

\subsection{The free energy}
\begin{figure}
  \centering
  \includegraphics[width=0.5\textwidth,keepaspectratio]{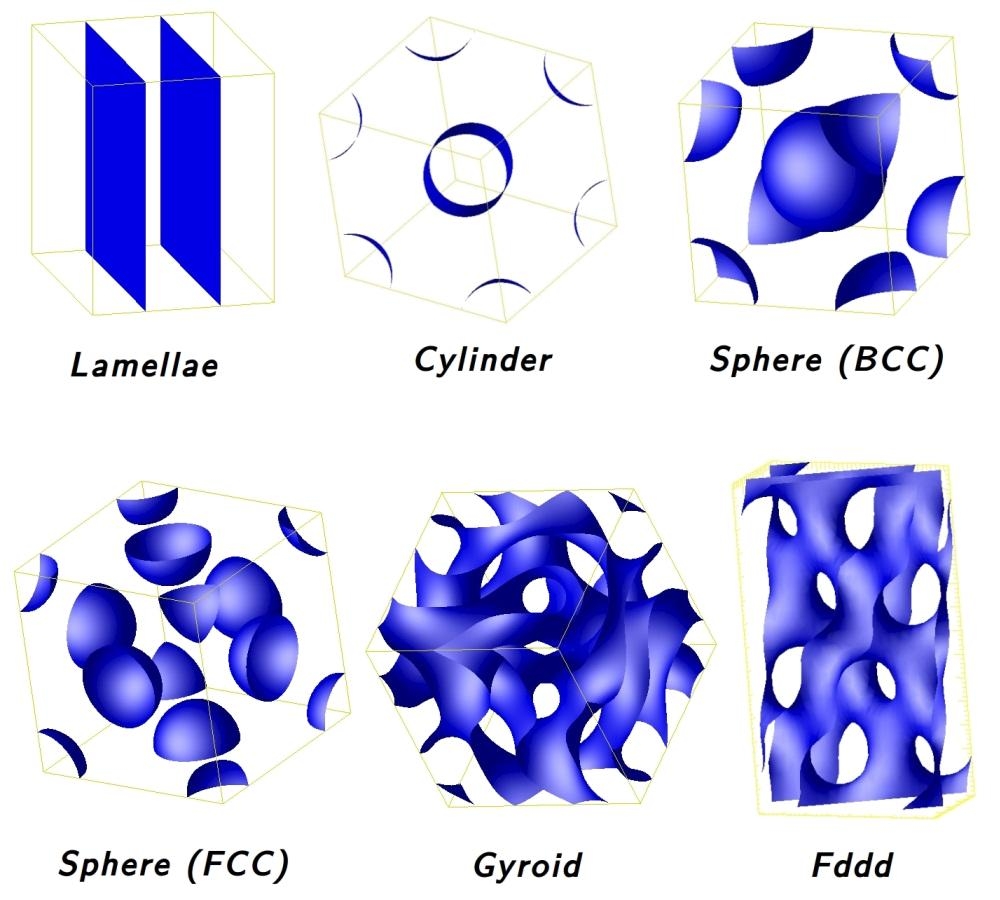}
  \caption{Bulk phases of Landau-Brazovskii model.}\label{phases}
\end{figure}
LB model was developed to treat fluctuation effects in systems undergoing weak
crystallization \cite{JChP87,PhRep1993}. This model can be viewed as a generic 
model for modulated phases occurring in a variety of physical and chemical 
systems, and its form is similar to many Landau-type free energy functionals 
for different kinds of materials. In addition, LB model can describe frequently 
observed patterns in all kinds of modulated systems, including lamellar(L), 
cylindral(C), spherical and gyroid(G) structures. 
Therefore our results could reveal properties of interfaces in a wide range of 
systems. 
In its scaled form, the LB free energy density is given by 
\begin{equation}\label{LB}
F=\frac{1}{V}\int_{\Omega}\md{\bm r}\{\frac{\xi^2}{2}[(\nabla^2+q_0^2)\phi]^2\}
+\frac{\tau}{2}\phi^2-\frac{\gamma}{3!}\phi^3+\frac{1}{4!}\phi^4, 
\end{equation}
where $q_0=1$ is the critical wavelength, $\xi,\ \tau,\ \gamma$ are 
phenomenological parameters, and $\phi$ is conserved, 
$$
\int\md\bm{r}\phi=0. 
$$ 
The parameters can be determined by measurable parameters in some cases. 
An example is the system of A-B diblock copolymer, 
in which these parameters are derived from $\chi N$ and $f$, 
where $\chi N$ is a normalized parameter characterizing the segregation of 
two blocks, and $f$ is the fraction of block A. 
The phases in LB model can be easily recognized by the isosurface of $\phi$, 
drawn in Fig. \ref{phases}. For the interfaces, we will also draw the 
isosurface to reveal their structures. 

\subsection{Boundary conditions}
Bulk values are needed for setting initial and boundary values of the problem. 
They are obtained by minimizing (\ref{LB}) with periodic boundary condition in 
all three directions. 
The existence of second derivatives in the energy functional requires $\phi$ 
and $\nabla \phi$ to be fixed at $x=\pm L$. These values can be easily computed 
with bulk profiles. 

When computing bulk profiles, the period lengths should be optimized as well.
This is because the size of the cell also affects the energy density of 
the system.
When two phases coexist, however, it is usually observed experimentally that 
two unit cells perturb a little from the optimal to match each other 
\cite{PRL1994, Ma94, JPCB2003, JACS2009}. 
To capture the interfacial structure in our framework, 
we slightly stretch the two bulks to let them have common period lengths. 
Then our framework for the interface could be applied directly. 
Such stretching is feasible since it will affect the energy within 
a negligibly small amount. 
It should be pointed out that besides the period matching, different 
ordered structures have certain preferences in orientation when they coexist, 
namely the rotations $\mathcal{R}_{\alpha},\ \mathcal{R}_{\beta}$ and the shifts 
${\bm b}_{\alpha},\ {\bm b}_{\beta}$ prefer certain values. 
Such epitaxial relationships are also noted in the experimental works 
mentioned above, and are studied in \cite{SoftM2011} extensively. 
We will examine such epitaxies as well as less optimal matching cases 
for the interfacial systems. 

The profiles of L, C and G are calculated with the common period 
$2\sqrt{6}\pi\times 2\sqrt{6}\pi\times 2\sqrt{6}\pi$, which is 
almost accurate for lamellar and cylinder, while $4\%$ smaller for gyroid. 
The number of meshes used in a unit cell is $32\times 32\times 32$. 
Details of discretization and optimization method are given in Appendix, 
in which acceleration techniques are included. 

\subsection{C-G and L-G interfaces}
\begin{figure}
  \centering
  \includegraphics[width=0.8\textwidth,keepaspectratio]{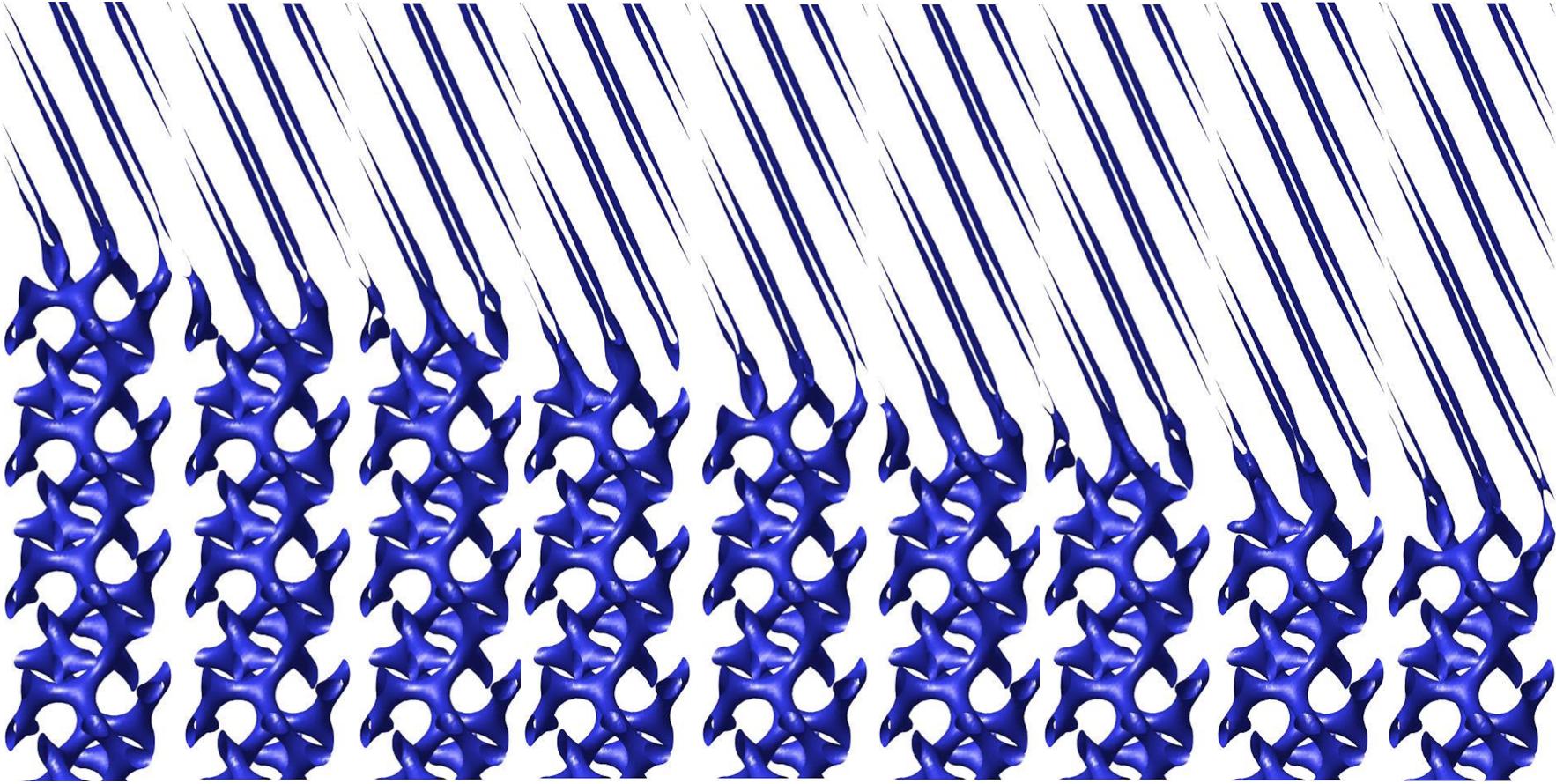}\\\vspace{10pt}
  \includegraphics[width=0.8\textwidth,keepaspectratio]{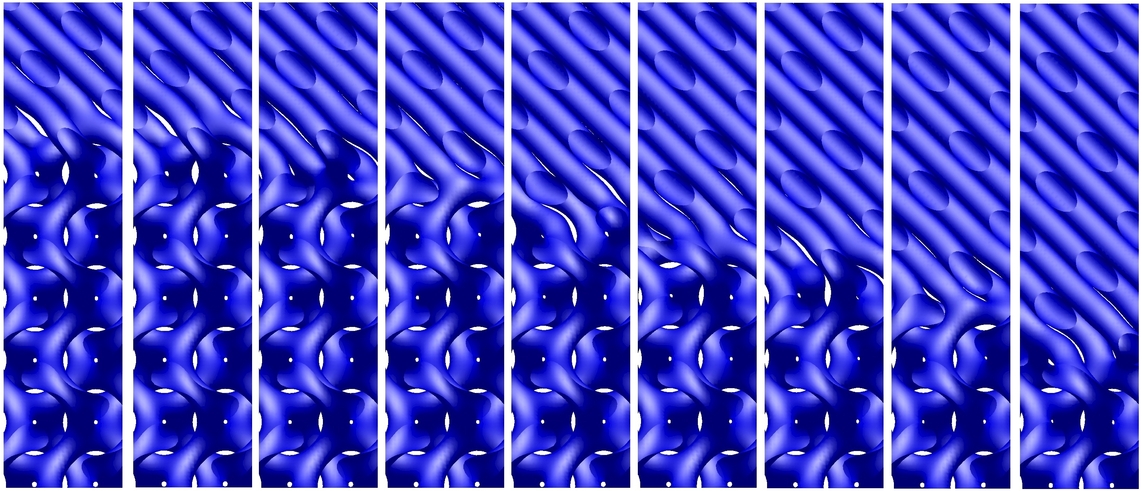}
  \caption{L-G and C-G interfaces at local minima (within two periods): 
    the epitaxially matching case. }
  \label{match}
\end{figure}
\label{numres}
We start from the cases in which two phases are epitaxially matched. 
In the lattice of G, the layer of L parallels to the plane $(11\bar{2})$ and 
the hexagonal lattice of C lies in the plane $(111)$. 
The results in Fig. \ref{match} are calculated with 
$\xi^2=0.0389,\ \gamma=0.0681$; $\tau=-0.0121$ for C-G interfaces 
and $\tau=-0.0159$ for L-G interfaces. 
Fig. \ref{match} presents C-G and L-G interfaces at local energy minima. 
Both C-G and L-G interfaces show four local minima within one period, 
and when moving a full period, the interfacial structures reappear. 
We are also able to catch how the interface moves from 
the ones at discrete locations. 

\begin{figure}
  \centering
  \includegraphics[width=0.48\textwidth,keepaspectratio]{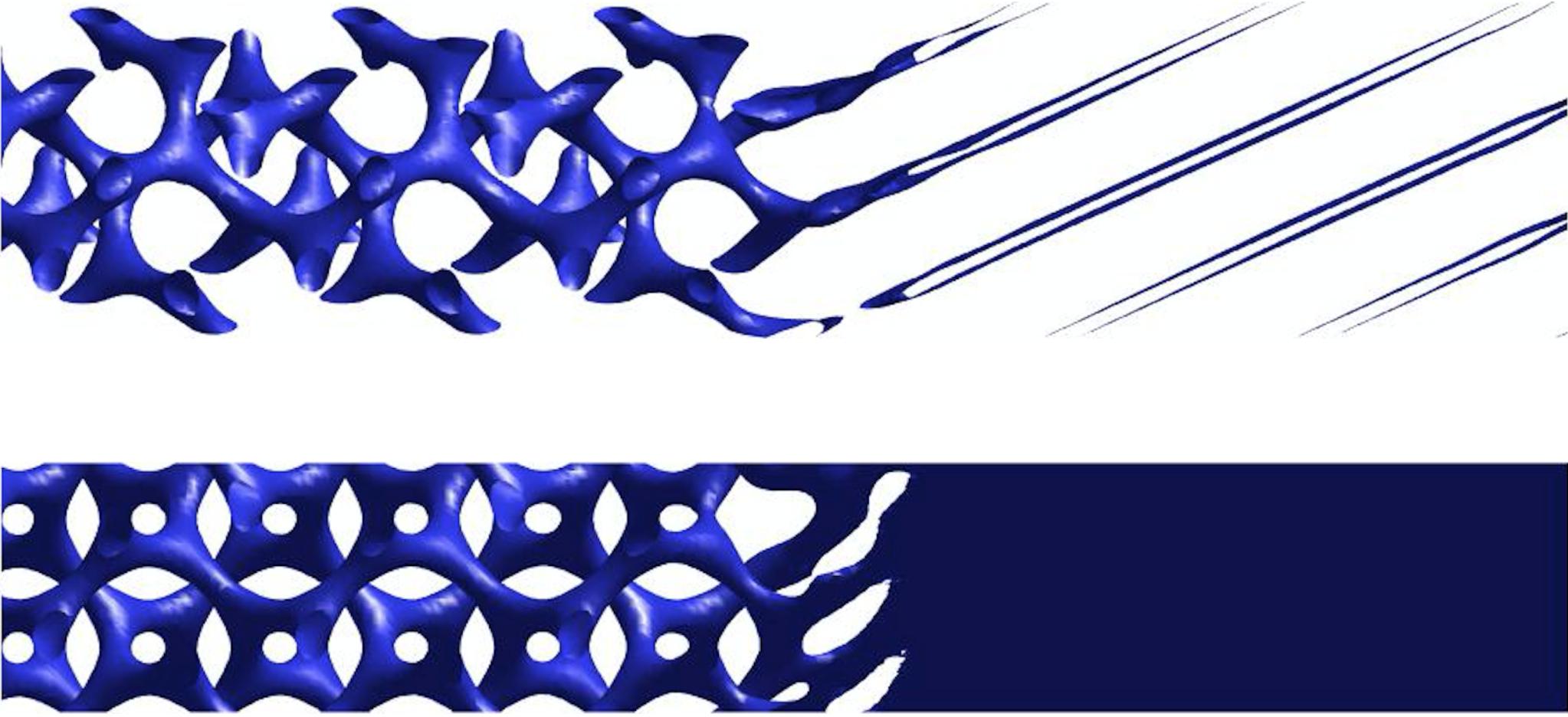}
  \includegraphics[width=0.48\textwidth,keepaspectratio]{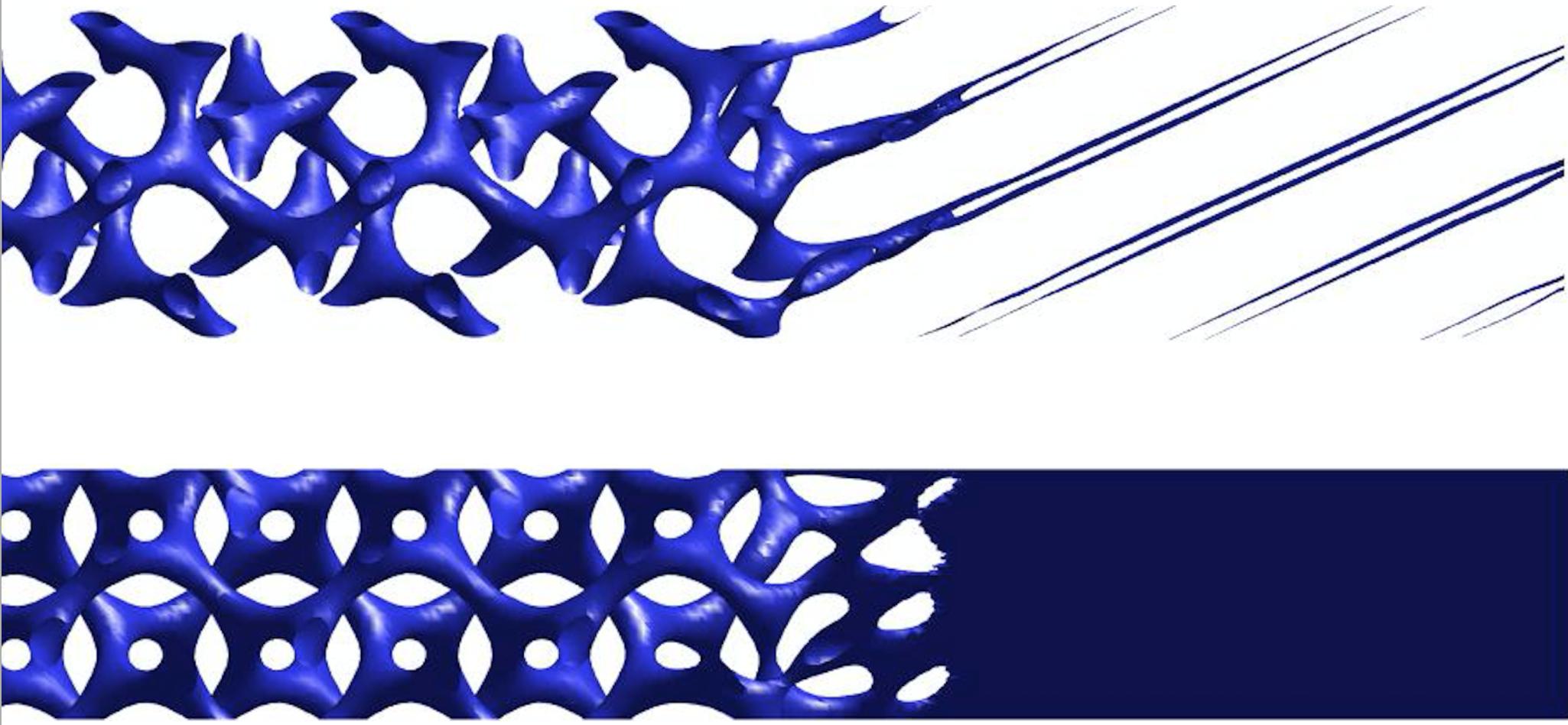}
  \caption{The L-G interface. Left: the epitaxially matching case; 
    Right: L is shifted a half period. Viewed along the layer of L (upper) 
    and the unit cell of G (lower). }
  \label{LGshift}
\end{figure}
\begin{figure}
  \centering
  \includegraphics[width=0.55\textwidth,keepaspectratio]{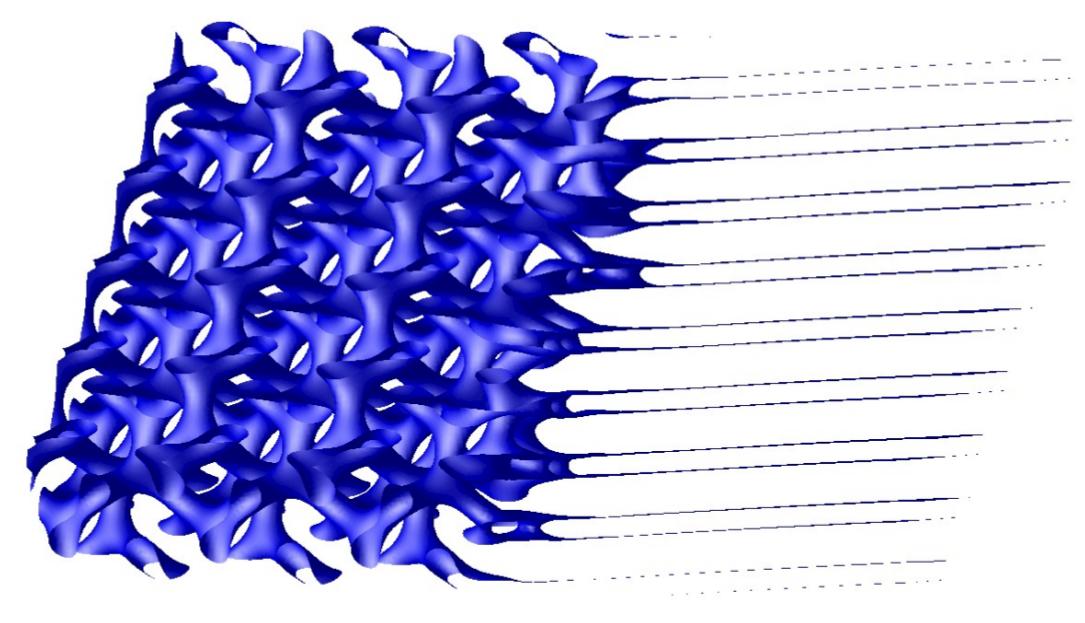}
  \includegraphics[width=0.4\textwidth,keepaspectratio]{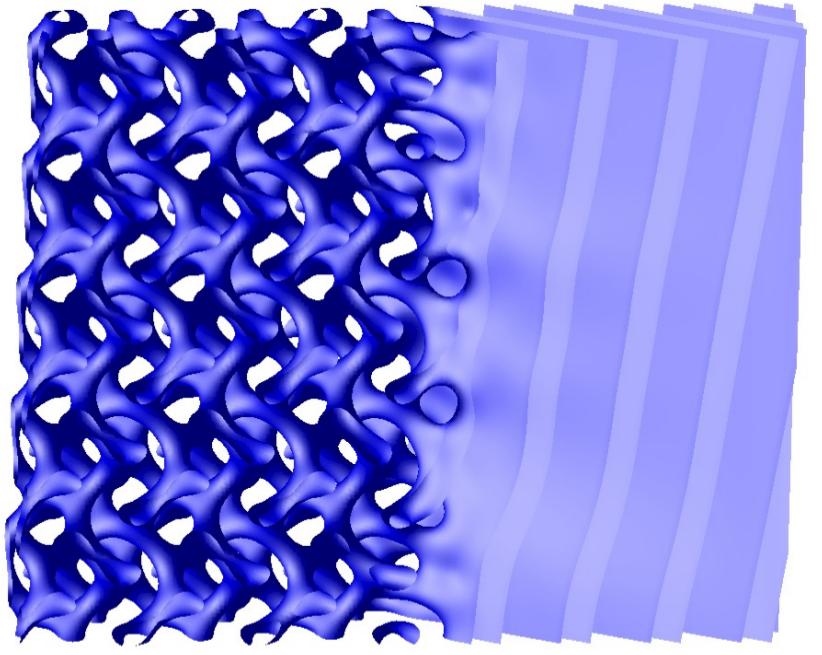}
  \caption{L-G interfaces with L rotated counter-clockwise 
    $\theta=\arcsin(3/5)$ (left), and $\theta+\pi/2$ (right). }
  \label{LGrotate}
\end{figure}
In the following, we will present some results of the non-matching cases, 
in which we can observe some interesting phenomena. 
The results described in this paragraph are calculated with 
$\xi^2=0.0389$, $\gamma=0.0601$, $\tau=-0.0121$. 
First, we examine the case where L is shifted a half period 
(Fig. \ref{LGshift} right). 
Comparing it with the epitaxially matching case (Fig. \ref{LGshift} left), 
we find the structure distinct from L and G in the middle. 
It resembles the metastable perforated layer structure, 
supporting the prevalent observation of perforated layer phase in the 
L$\leftrightarrow$G transitions (see the discussion in \cite{PRL_Ncl}). 
Next we look at the effects of relative rotation. 
In Fig. \ref{LGrotate} L is rotated $\theta=\arcsin(3/5)$ and 
$\theta+\pi/2$ counter-clockwise respectively, G unchanged. 
Local distortions help to keep their connection, leading to non-planar 
interfaces. 

\begin{figure}
  \centering
  \includegraphics[width=0.48\textwidth,keepaspectratio]{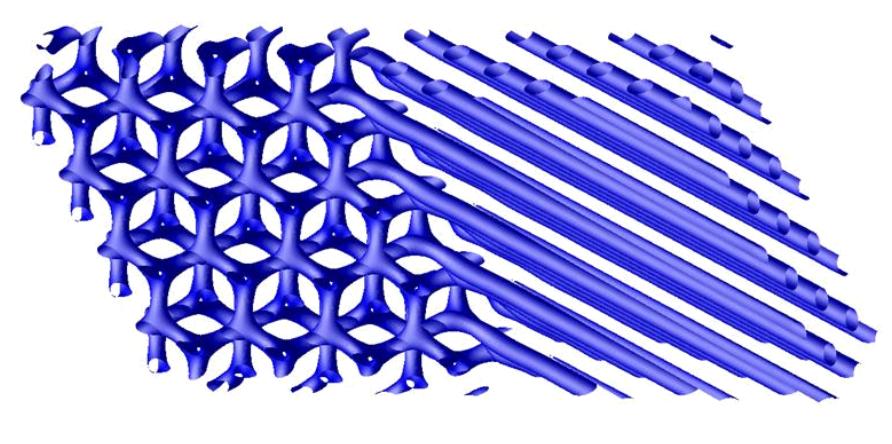}
  \includegraphics[width=0.45\textwidth,keepaspectratio]{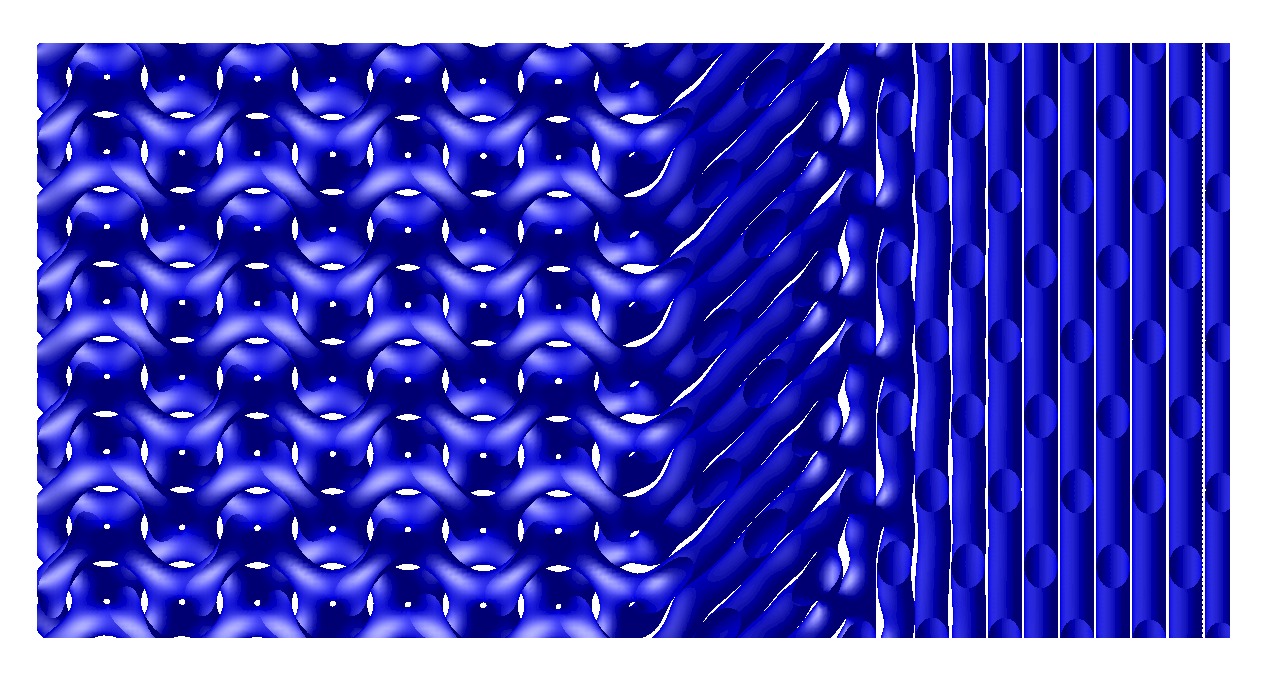}
  \caption{Left: C-G interface with newly found epitaxy $(1\bar{1}0)$. 
    Right: C is rotated $\pi/2$ from the new epitaxy and  
   cylinder along $\left<111\right>$ direction is found in between.}
  \label{CGnew}
\end{figure}

The next pair of examples are based on the newly-found epitaxially 
relationship between C and G \cite{JACS2009}, where the lattice of C is 
slightly deformed from the regular hexagon and lies in the plane $(1\bar{1}0)$. 
The parameters are chosen as $\xi^2=0.0375$, $\gamma=0.0757$, $\tau=-0.0102$. 
The interface is drawn in the left of Fig. \ref{CGnew}, 
which is planar with smooth connection. 
The right of Fig. \ref{CGnew} shows the interface 
where C is rotated $\pi/2$ in the $x$-$y$ plane. 
To connect two phases, C of the regular hexagonal type with classical epitaxy 
appears in between. 

\section{Summary\label{concl}}
In this paper, we propose a general framework for computing interfacial 
structures of ordered phases with optimization approach. 
In the framework boundary conditions imposed are compatible with 
both bulk phases. 
Unlike computing interface between disordered phases where constraints need to 
be introduced, the problem is well-posed because anistropy leads to local 
minima that stabilize the location of interface. 
As an example, we apply the framework to Landau-Brazovskii model. 
It works efficiently in both epitaxially matching and non-matching cases. 

So far we show the application of the framework to special cases where public 
period can be found in the $y$-$z$ plane. The choice of basis functions 
(\ref{basis}) actually allows us to investigate quasiperiodic phases. 
Thus in the future we aim to apply this framework to broader cases, 
especially for quasiperiodic phases. 
\vspace{20pt}

\textbf{Acknowldegment}
Pingwen Zhang is partly supported by National Natural Science Foundation of 
China (Grant No. 11421101, No. 11421110001 and No. 21274005).

\appendix
\section{Numerical details}
We describe the numerical details of discretization and optimization. 
For the discretization of density profile, 
finite difference scheme is adopted in the $x$-direction. In the $y$-$z$ plane, 
both finite difference scheme and Fourier expansion can be used. 

In the $x$-direction, Laplacian is approximated by
$$
\partial_x^2\phi(x_k)\approx\delta_x^2\phi(x_k)=
\frac{\phi_{k+1}-2\phi_k+\phi_{k-1}}{\Delta x^2}. 
$$
The same approximation is adopted when using finite difference scheme in 
the $y$-$z$ plane. 
When using Fourier expansion in the $y$-$z$ plane, we write $\phi$ as 
$$
\phi(x,y,z)=\sum_{\bm G} \phi_{\bm G}(x)\exp(i({\bm G}\cdot{\bm r}')). 
$$
where ${\bm G}=m{\bm b}_1'+n{\bm b}_2',\quad |m|,|n|\le N$, 
$\bm{b}_i'$ are reciprocal vectors with respect to the lattice in the $y$-$z$ 
plane, and ${\bm r}'=(y,z)$. Note that $\phi$ is real-valued, thus
it requires $\phi_{-{\bm G}}(x)=\phi^*_{\bm G}(x)$. 

The anchoring boundary conditions at $x=\pm L$ are implemented by adding two 
extra grids on each side and setting $\phi$ equal to bulk values. This can be 
equivalently viewed as approximating boundary conditions with finite difference 
scheme, 
$$
\phi(-L)=\phi_{\alpha}(-L), \qquad 
\partial_x\phi(-L)\approx\frac{\phi_{0}-\phi_{-1}}{\Delta x}
=\frac{\phi_{\alpha}(x_{0})-\phi_{\alpha}(x_{-1})}{\Delta x}
\approx\partial_x\phi_{\alpha}(-L).
$$

The gradient vector can be calculated by 
$$
\nabla \tilde{F}(\phi(x))=\xi^2[(\delta_x^2+\delta_y^2+\delta_z^2+1)^2+\tau]\phi(x) 
-\frac{\gamma}{2}\phi^2(x)+\frac{1}{6}\phi^3(x)
$$
for finite difference method, and by
\begin{equation}
\nabla \tilde{F}(\phi_{\bm G}(x))=\xi^2[(\delta_x^2-{\bm G}^2+1)^2+\tau]\phi_{\bm G}
-\frac{\gamma}{2}\sum_{{\bm G}_1+{\bm G}_2={\bm G}}\phi_{{\bm G}_1}
\phi_{{\bm G}_2}+\frac{1}{6}\sum_{{\bm G}_1+{\bm G}_2+{\bm G}_3={\bm G}}
\phi_{{\bm G}_1}\phi_{{\bm G}_2}\phi_{{\bm G}_3} \label{EL}
\end{equation}
for Fourier expansion. The convolution sum can be calculated by FFT. 

The conservation of $\phi$ is attained by a projection on the gradient vector: 
for finite difference scheme, we use 
$$
\nabla F(\phi(x))=\nabla\tilde{F}(\phi(x))-c; 
$$
and for Fourier expansion, we use 
$$
\nabla F(\phi_{\bm G}(x))=\nabla \tilde{F}(\phi_{\bm G}(x))-c\delta({\bm G}=0). 
$$
In the above, $c$ can be determined by the following observation: 
if we set $\phi=\phi_{\alpha}$ for $x<0$ and $\phi=\phi_{\beta}$ for $x>0$, 
the constraint is satisfied. So we can just require 
$$
\int_S\int_{-L}^L\phi \md x\md y\md z=
\int_S\md y\md z(\int_{-L}^0\phi_{\alpha}\md x+\int_0^L\phi_{\beta}\md x)
\triangleq c_0. 
$$

For finite difference scheme, we use a gradient method
$$
\phi^{n+1}(x)- \phi^n(x)=-a_n\nabla F(\phi^n(x)). 
$$
The coefficient $a_n$ is altered adaptively with Barzilai-Borwein method
\cite{BB0,BB}. 
For Fourier expansion we use a semi-implicit scheme to solve 
the Euler-Langrange equation (\ref{EL}) with 
\begin{eqnarray*}
\frac{\phi^{n+1}_{\bm G}(x)-\phi^{n}_{\bm G}(x)}{\Delta t}&=&
-\xi^2[(\delta_x^2-{\bm G}^2+1)^2+\tau]\phi^{n+1}_{\bm G}\\
&&+\frac{\gamma}{2}\sum_{{\bm G}_1+{\bm G}_2={\bm G}}\phi^n_{{\bm G}_1}
\phi^n_{{\bm G}_2}-\frac{1}{6}\sum_{{\bm G}_1+{\bm G}_2+{\bm G}_3={\bm G}}
\phi^n_{{\bm G}_1}\phi^n_{{\bm G}_2}\phi^n_{{\bm G}_3}\\
&&-c\delta({\bm G}=0).
\end{eqnarray*}

\bibliographystyle{unsrt}
\bibliography{alg_bib} 

\end{document}